\documentclass[twocolumn,pra,aps,raggedbottom, superscriptaddress]{revtex4-2}
\usepackage[table,xcdraw]{xcolor}
\usepackage{amsmath,amsfonts,amssymb,graphicx,hyperref,tikz,natbib,array}
\usepackage{mathptmx,textcomp,braket,multirow}
\usepackage[T1]{fontenc}
\usepackage[utf8]{inputenc}
\usepackage{multirow}
\hypersetup{colorlinks,citecolor=blue,filecolor=black,linkcolor=blue,urlcolor=blue}
  
\definecolor{lime}{HTML}{A6CE39}
\DeclareRobustCommand{\orcidicon}
{
	\begin{tikzpicture} 
	\draw[lime, fill=lime] (0,0) circle [radius=0.15] node[white] {{\fontfamily{qag}\selectfont \tiny ID}};
	\draw[white, fill=white] (-0.0625,0.095) 	circle [radius=0.007];
	\end{tikzpicture}
	\hspace{-2.2mm}
}
\newcommand\orcidID[1]{\href{https://orcid.org/#1}{\orcidicon}}

\newcommand{\be}{\begin {equation}}
\newcommand{\ee}{\end {equation}}
\newcommand{\beqa}{\begin {eqnarray}}
\newcommand{\eeqa}{\end {eqnarray}}
\newcommand{\mb}{\mathbf}
\newcommand{\Sch}{Schr\"odinger }
\newcommand{\Exp}[1]{\text{e}^{#1}}

\begin{document}

\title{Phase-Controlled Ramsey Interference of XUV Photoelectrons}

\author{Neha Kukreti} 
%\email{nehakukreti1111@gmail.com}
\author{Amol R. Holkundkar\orcidID{0000-0003-3889-0910}}
\email{amol@holkundkar.in}  
\affiliation{Department of Physics, Birla Institute of Technology and Science - Pilani, Rajasthan, 333031, India.}
 
\date{\today}

\begin{abstract}

We investigate Ramsey-type quantum interference in photoelectron momentum distributions generated by two time-delayed, linearly polarized extreme-ultraviolet (XUV) laser pulses. The electron dynamics are studied by solving the full-dimensional time-dependent Schr\"odinger equation within the single-active-electron approximation for neon initially prepared in a current-carrying $2p_+$ state. The coherent superposition of electron wave packets released by the two pulses gives rise to pronounced interference fringes in both energy-resolved spectra and angle-resolved momentum distributions. We demonstrate that the fringe positions are governed by a Ramsey phase accumulated during the interpulse delay, resulting in a linear dependence on the relative carrier-envelope phase and an inverse scaling of the fringe spacing with the delay time. By systematically varying the laser intensity, we establish that the observed modulations originate from temporal quantum interference rather than Autler--Townes splitting. Analysis of the time-resolved bound-state population dynamics reveals that carrier-envelope-phase dependent bound--bound coupling dominated by transient population transfer to the $2s$ state, which controls the interference contrast. The accumulated phase is further interpreted in terms of a dynamic Stark shift of the dressed bound states, which is quantitatively reproduced using a reduced two-level model.

\end{abstract}

\maketitle

\section{Introduction}

The advent of attosecond science \cite{Corkum2007Attosecond} has opened new frontiers for probing and controlling ultrafast electron dynamics in atoms and molecules \cite{Nisoli2017Attosecond}. In particular, the development of extreme-ultraviolet (XUV) and soft-x-ray light sources based on high-order harmonic generation (HHG) and free-electron lasers (FELs) has profoundly transformed the study of ultrafast electronic dynamics in atoms, molecules, and solids \cite{Calegari2021PRL,Emma_2004,Heslar2017_PRA,Mandal2022_OptExp,NEFEDOVA20179}. XUV pulses enable direct access to electron motion on its intrinsic timescale, allowing high-precision investigations of ultrafast photoionization \cite{PhysRevA.107.023101,PhysRevA.111.013110}, quantum coherence, and wave-packet evolution~\cite{Calegari_2016, PhysRevLett.71.1994, PhysRevLett.130.083202}.

A major current trend in ultrafast science is the exploitation of light polarization and phase as control parameters for encoding and retrieving dynamical information \cite{PhysRevA.111.L041101,PhysRevResearch.7.023268}. Circular and elliptical polarizations, in particular, have enabled detailed studies of photoelectron circular dichroism \cite{McKay2025AICD,PhysRevLett.107.253002,miloevi_2015}, revealing sensitivity to orbital angular momentum, electronic chirality \cite{Beaulieu2017_Science,Neufeld2022_OptExp,PhysRevA.111.033118}, and circular dichroism from non-chiral nanostructures \cite{ZambranaPuyalto2014CD,sercel2020circular}. These phenomena arise from the interplay between light helicity \cite{Ke2021Helicity,PhysRevA.111.043519} and the intrinsic angular momentum of the electronic wave function, making polarization- and phase-controlled photoionization powerful probes of symmetry and ultrafast dynamics \cite{Galan2017_OptExp,Reich2016_PRL}. Consequently, understanding how the phase, polarization, and temporal structure of driving fields influence photoelectron emission has become a topic of considerable interest \cite{PhysRevA.111.023112, PhysRevLett.134.203201,PhysRevA.111.053115,PhysRevLett.134.073201,PhysRevA.111.013106,PhysRevResearch.7.023034}.

Photoionization has long served as a cornerstone for probing electronic structure and dynamics in atomic, molecular, and optical physics. Early angle-resolved photoelectron spectroscopy established the connection between photoelectron angular distributions and the symmetry of the initial bound state. With the advent of ultrafast laser sources, time-resolved photoelectron spectroscopy enabled direct tracking of wave-packet evolution, coherence, and population transfer during laser–matter interaction \cite{PhysRevResearch.7.023268,PhysRevA.107.023101,PhysRevLett.130.083202}. In pump–probe schemes, ionization acts as a projective measurement, converting transient quantum coherence into measurable momentum- and energy-resolved observables \cite{ramasesha2016real}. As a result, photoelectron momentum distributions (PMDs) have become a sensitive tool for disentangling competing ionization pathways, probing phase accumulation, and identifying the role of intermediate bound and continuum states \cite{PhysRevLett.134.163201,PhysRevA.111.L041102,gebre_2024}.

Among interference-based phenomena in photoionization, Ramsey-type interference occupies a distinct position \cite{Ramsey1950PR,Fang2023_RAMSEY_CPT,PhysRevA.102.013511}. Originally developed in the context of atomic clocks and precision spectroscopy, Ramsey interference arises from the coherent superposition of quantum amplitudes generated by two temporally separated excitation events \cite{ma2024attosecond}. In photoionization, this manifests as energy- and angle-dependent interference fringes whose spacing is determined by the interpulse delay and whose phase is controlled by the relative optical phase between the pulses. This mechanism is fundamentally different from Autler–Townes splitting, which originates from strong-field dressing of bound states and leads to intensity-dependent spectral doublets associated with Rabi oscillations \cite{yang2025,Bayer2023PRA}. While Autler–Townes spectra reflect quasistationary dressed-state formation, Ramsey interference is a purely temporal effect governed by phase accumulation \cite{buth2015ramsey} between distinct ionization pathways. Disentangling these two mechanisms is therefore essential for the correct interpretation of interference structures observed in ultrafast photoelectron spectra.

In the XUV regime, Ramsey interference has received comparatively limited attention, particularly in the single-photon ionization limit where strong-field effects such as ponderomotive shifts and multiphoton couplings are negligible. Moreover, the role of the carrier-envelope phase (CEP) in controlling Ramsey interference in XUV photoionization remains largely unexplored. At XUV frequencies, CEP effects are often assumed to be weak for single-pulse ionization due to the short optical cycle. However, when ionization pathways are temporally separated, the CEP directly enters the accumulated phase, thereby strongly influencing the interference contrast and fringe structure. At the same time, laser-induced phase shifts raise important questions regarding the possible role of dynamic Stark effects \cite{PhysRevA.109.L020801} in shaping Ramsey interference in the continuum.

In this work, we address these issues by performing a full-dimensional time-dependent Schr\"odinger equation (TDSE) study of XUV photoionization driven by two time-delayed, linearly polarized pulses. We focus on neon initially prepared in a current-carrying $2p_{+}$ state, which provides a natural sensitivity to angular momentum and phase. By systematically varying the carrier-envelope phase (CEP), interpulse delay, wavelength, and laser intensity, we demonstrate that the observed interference patterns in the photoelectron momentum and energy distributions originate from Ramsey-type temporal interference rather than Autler--Townes splitting. Time-resolved population analysis reveals that CEP-dependent bound--bound coupling, dominated by transient population transfer to the $2s$ state within the single-active-electron approximation, plays a central role in controlling the accumulated phase. We further show that
the time dependent energy shift of the bound-state is quantitatively captured by a reduced two-level model, providing a transparent physical interpretation of the full TDSE results.

From an experimental perspective, the present study is directly relevant to ongoing efforts in ultrafast spectroscopy \cite{Peng2019}. Advances in HHG-based XUV sources and free-electron lasers have enabled the generation of phase-stable, time-delayed XUV pulse pairs with controlled polarization and relative phase \cite{Calegari2021PRL}. CEP tagging and stabilization techniques, together with momentum-resolved electron detection methods such as velocity-map imaging \cite{Kheifets2021_VMI_XUV} and COLTRIMS \cite{Weger_13,Kunitski2019_COLTRIMS,Hickstein2012_CEP_PMD}, make the observation of CEP-controlled Ramsey interference experimentally feasible. Our results therefore provide clear theoretical benchmarks and physical guidance for future experiments aimed at exploiting temporal interference as a phase-sensitive probe of electronic dynamics in the XUV regime.

The paper is organized as follows. Section~\ref{sec:theory} presents the theoretical framework and numerical methodology based on solving the TDSE within the SAE approximation. Section~\ref{sec:results} presents and discusses the results which are divided into five subsections. Finally, the main conclusions and outlines perspectives for future studies are summarized in Sec. ~\ref{sec:conclusion}. Moreover, calculations related to the dressed-energy level is discussed in Appendix~\ref{app:2s}.

\section{Theoretical and Numerical Considerations} \label{sec:theory}
 
We have developed a full dimensional time-dependent \Sch equation (TDSE) solver within the  single-active-electron (SAE) approximation using the time-dependent generalized pseudospectral (TDGPS) method \cite{TONG1997119,Tong_2017,Holkundkar2023_PhysLettA, Rajpoot2023_JPhysB}. The TDSE in the length gauge is written as:
\be
i \frac{\partial}{\partial t} \ket{\psi(\mb{r},t)} = [H_\text{0} + H_\text{L}(t)] \ket{\psi(\mb{r},t)},
\label{tdse0}
\ee
   
where, $H_\text{0} = -\nabla^2/2 + V(r)$ is the field free Hamiltonian and $H_\text{L}(t) = -\mb{r}\cdot\mb{E}(t)$ is the interaction Hamiltonian in the length gauge, with $\mb{E}(t)$ being the temporal profile of driving laser field within the dipole approximation and given by the superposition of two time delayed linearly polarized pulses, $\mb{E}(t) = \mb{E_1}(t,\phi_1) + \mb{E_2}(t-\tau,\phi_2)$ with, \be\mb{E_j}(t,\phi_j) = E_0 \sin^2(\pi t/T) \sin(\omega_0 t + \phi_j)\  \mb{\hat{e}_j} \ee here, $E_0$ [a.u.] $\simeq 5.342\times 10^{-9}\sqrt{I_0}$ is the field amplitude, with the peak intensity $I_0$ expressed in W/cm$^2$, and $\omega_0 \simeq 1.14\ \text{a.u.}$ is the carrier frequency of the laser corresponding to the XUV wavelength $\lambda = 40$ nm. Each pulse is nonzero only within the temporal window $0 \le t \le T$, and the second pulse envelope is shifted in time by the delay $\tau$. Both pulses are polarized along the $x$ axis, i.e., $\hat{\mathbf{e}}_1 = \hat{\mathbf{x}}$ and $\hat{\mathbf{e}}_2 = \hat{\mathbf{x}}$. $\phi_j$ denote the carrier-envelope phases (CEPs) of the two pulses, however throughout the manuscript $\phi_1 = 0$ is considered and $\phi = \phi_2 - \phi_1$ will hereafter be referred to as the CEP of the pulse (relative phase between the two pulses). Pulse duration of each pulse is considered to be $T = 5 \tau_0$, with $\tau_0 \text{[a.u.]} = 2\pi/\omega_0$ being the duration of the one cycle corresponding to respective wavelength. All laser parameters considered here are experimentally accessible with current laser technology  \cite{PhysRevLett.131.045001, Huang2018Polarization}. Atomic units ($|e| = m_e = \hbar = 1$) are used throughout unless otherwise stated.

The atomic Coulomb potential $V(r)$ in the $H_\text{0}$ is modeled under single active electron (SAE) approximation by an empirical expression given by \cite{Tong_2005}:
\be V(r) = -\frac{1 + a_1\ \Exp{-a_2 r} + a_3\ r\ \Exp{-a_4 r} + a_5\ \Exp{-a_6 r}}{r} \label{potential}\ee
The values of the coefficients $a_i$'s for atomic species Hydrogen, Helium, Neon and Argon are tabulated in Ref. \cite{Tong_2005}. This empirical expression of $V(r)$ is based on the self-interaction-free density functional theory within the SAE approximation. 
An advantage of combining the TDGPS method with this atomic potential is that no soft-core regularization is required, in contrast to Cartesian-grid implementations.
 
In TDGPS, the radial domain $r = [0,R_{max}]$ is mapped on the range $s =  [-1,1]$ which is further discretized using the \textit{roots of the Legendre polynomial}. As a result no matter how large the radial points we consider between $s \equiv [-1,1]$ the radial point $r = 0$ corresponding to $s = -1$ is never incorporated in the simulation domain \cite{TONG1997119,Tong_2017,Wang1994_PhysRevA}. This accurate consideration of the model potential enables us to calculate the ionization potential and other quantities with great precision. For example, in this work we adopted the radial simulation domain of $R_{\text{max}} = 500$ atomic units (a.u.), with the last $50$ a.u. utilized as a masking region to absorb the outgoing wavefunction.

The eigen-energies \(E_{n\ell}\) and their corresponding radial eigenfunctions \(R_{n\ell}(r)\) of the field-free Hamiltonian \(H_0\) for each partial wave \(\ell\) are evaluated before the propagation using the split-operator-method \cite{Tong_2017}.  Since the radial eigenstates along with the spherical harmonics $\{ R_{n\ell}(r)\, Y_{\ell m}(\theta,\varphi) \}$ form a complete basis set within the finite simulation box, the time-dependent wavefunction is expanded in the energy basis of $H_0$ as:
\be
\psi(\mb{r},t) = \sum_{n,\ell,m} C_{n,\ell, m}(t)\, R_{n\ell}(r)\, Y_{\ell m}(\theta,\varphi).
\label{psi_t}
\ee
where, $C_{n\ell m}(t) $ is the time-dependent probability amplitude of the corresponding state. These coefficients enable us to keep track of the time resolved population dynamics during the interaction. Since the dynamics are dominated by single-photon ionization, and hence the results are found to be converged with $\ell_\text{max} = 15$. Moreover, the TDGPS method enables to use coarser time steps without affecting the results, in our case the results are found to be converged for simulation time step $\delta t = 0.2$ a.u. 
  
In order to  obtain the converged electron energy distribution, we have propagated the wavefunction for additional 50 optical cycles after the second pulse ended ($T_\text{end,II}$) i.e. $T_\text{final} = T_\text{end,II} + 50 \tau_0$.  The final wavefunction say $\psi_\text{final}(\mb{r},T_\text{final})$ is masked for $r \leq r_0$  with $r_0 = 100$ a.u. to obtain the continuum part of the wavefunction, $\psi_\text{cont}(\mb{r}) = \mathrm{M}(r,r_0) \psi_\text{final}(\mb{r},T_\text{final}) $, where, $\mathrm{M}(r,r_0) = [1 + \text{e}^{-3 (r - r_0)}]^{-1}$.  This masking would prevent the contamination from the bound-state component and from the region where the SAE potential deviates from a hydrogenic Coulomb potential. The PMD is then obtained by projecting the $\psi_\text{cont}$ on the analytically known hydrogenic Coulomb Scattering state  \cite{PhysRevA.76.063407}.   Although the SAE potential for Ne differs from a pure Coulomb form in the inner region, its asymptotic behavior is identical to $-1/r$. Once the electron wavepacket reaches $r\gtrsim 100$ a.u., where the projection is performed, the difference between the exact SAE continuum states and the analytic hydrogenic Coulomb scattering states reduces to a short--range  phase that does not affect the asymptotic momentum distribution \cite{PhysRevA.76.063407}.  Therefore, projecting onto Coulomb scattering states is a theoretically sound and computationally efficient approximation for differential ionization probability   $dP/d\mathbf{k}$ for asymptotic momentum $\mb{k}$,
\begin{equation}
\frac{dP}{d\mb{k}} = | \braket{\psi_{k}^{-,C} | \psi_\text{cont}(\mathbf{r})} |^2,
\end{equation}
where, $\ket{\psi_k^{-,C}}$ is the analytically known Coulombic waves \cite{PhysRevA.76.063407}. These PMD are also benchmarked with another method as discussed in Refs. \cite{murakami_2013, murakami_2020}, though they are computationally expensive, where at each time step the wavefunction need to be filtered out and after the simulation all the contributions need to be coherently added and then the PMD is obtained by projecting on the Volkov states. 
 
\section{Results and Discussion} \label{sec:results}

\subsection{Understanding PMD} \label{subsec:pmd}
To begin, we examine the interaction of two time-delayed, linearly polarized XUV pulses (both polarized along the $x$--axis), each of duration 5 optical cycles and separated by a delay $\tau = 6\tau_0$, with peak intensity $I_0 = 10^{14}$~W/cm$^2$, interacting with neon prepared in the current-carrying $2p_{+}$ bound state. This state corresponds to the magnetic sublevel $m = +1$ of the $2p$ manifold and carries a well-defined azimuthal probability current associated with its orbital angular momentum. The field-free energy of this state in our single--active--electron (SAE) potential is $E_{0} = -0.7933\ \text{a.u.}$ 
 
\begin{figure}[t]
\centering
\includegraphics[width=0.8\linewidth]{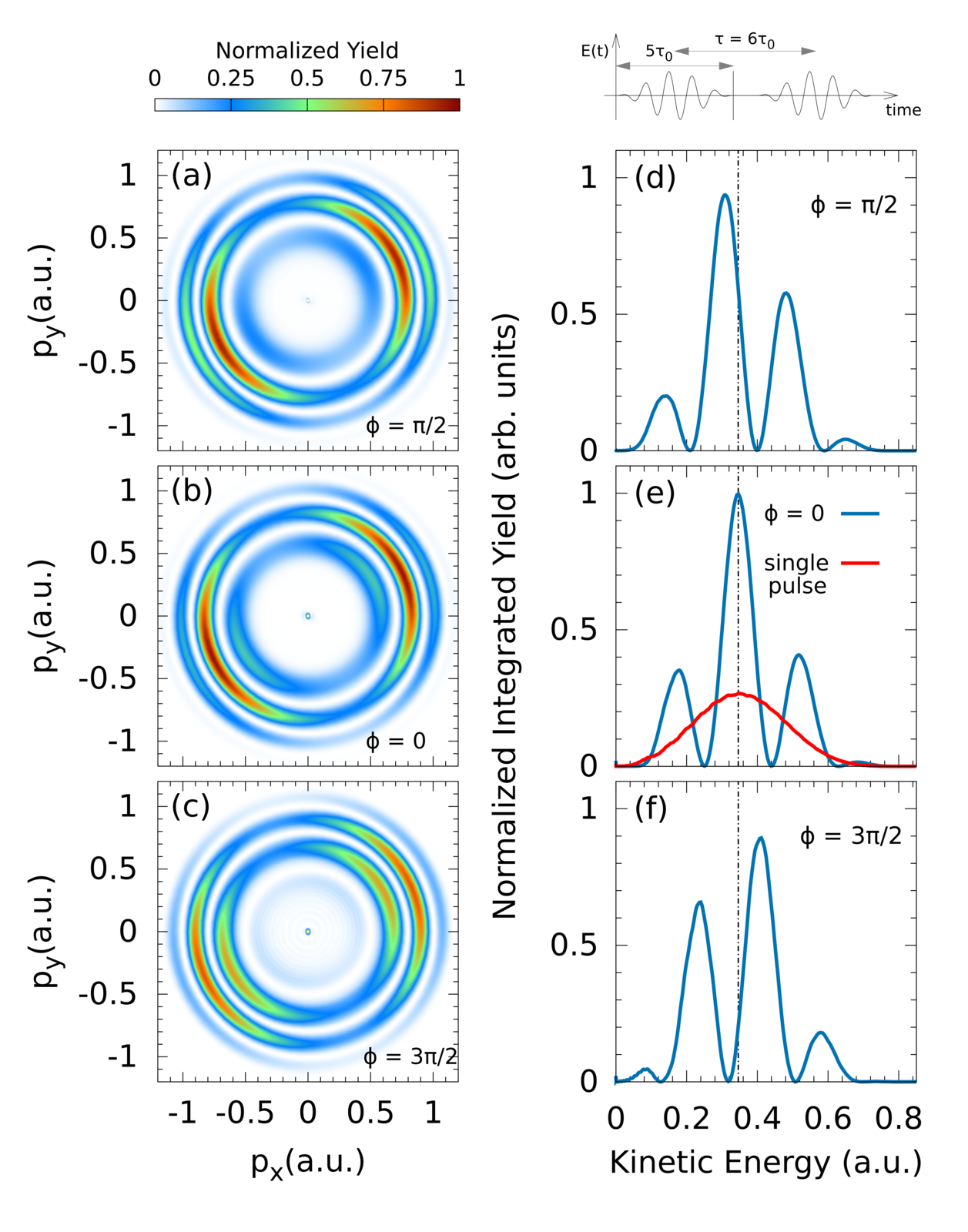}
\caption{Photoelectron momentum distribution in $x-y$ plane is shown for CEP $\phi = \pi/2$ (a), $\phi = 0$ (b) and $\phi = 3\pi/2$ (c). Total angle-integrated energy distribution for these respective cases are illustrated in (d), (e) and (f), along with the case when only single pulse is used [red curve in panel (e)]. Yield is normalized with respect to the $\phi = 0$ case for all the results in this figure. The schematic diagram of the two 5 cycle pulses, separated by the delay $\tau = 6\tau_0$ is also shown. Vertical dashed line in (d), (e) and (f) represents the $E_{\mathrm{kin}}\simeq 0.346$ a.u. [Eq. \eqref{ener_kin}]. }
\label{fig1}
\end{figure} 

We present the PMDs for three different relative CEP values in Fig.~\ref{fig1}(a)–(c), with the corresponding angle-integrated energy spectra shown in Fig.~\ref{fig1}(d)–(f). The momentum distributions exhibit a sequence of concentric rings whose radial intensity modulations depend sensitively on the CEP. These structures arise from the coherent superposition of electron wave packets released by the first and second pulses respectively. The time delay $\tau = 6\tau_{0}$ allows the wave packet launched by the first pulse to accumulate a CEP-dependent phase before the arrival of the second pulse, and this phase difference is directly mapped onto the angular structure of the PMD.

For the reference case $\phi = 0$ [Fig.~\ref{fig1}(b)], the yield is distributed among multiple concentric rings with a specific intensity balance relative to the central peak. Changing the CEP to $\phi = \pi/2$ or $\phi = 3\pi/2$ does not alter the angular structure of the PMD; instead,  redistributes the relative brightness of these rings [Figs.~\ref{fig1}(a) and \ref{fig1}(c)]. This behavior reflects the fact that variations of the CEP shift the electric-field phase within each pulse and thereby modify the interference between the ionization amplitudes associated with the two temporally separated pulses.

The angle-integrated energy spectra in panels (d)–(f) further illustrate the role of temporal quantum interference. Each spectrum exhibits multiple peaks arising from the interference between electron wave packets emitted at different times, both within and between the two pulses. The CEP modifies the phase accumulated between these ionization pathways and therefore controls both the contrast and relative height of the spectral peaks. In contrast, the single-pulse spectrum [red curve in Fig.~\ref{fig1}(e)] displays a smooth and broad energy distribution without such modulations, demonstrating that the multi-peak structure is a direct consequence of two-pulse interference.

The single-pulse case provides as a natural reference for interpreting the two-pulse results. For a single XUV pulse, the dominant peak in the energy spectrum originates from single-photon ionization of the initial $2p_{+}$ state and is located at
\[
E_{\mathrm{kin}} = \hbar\omega_{0} - I_{p} - U_{p}.
\]
At the XUV photon energies employed here, the ponderomotive energy is negligibly small
($U_{p}\sim5.5\times10^{-4}$~a.u. for $I_{0}=10^{14}$~W/cm$^{2}$) and can therefore be neglected. The peak position
is thus well approximated by
\be
E_{\mathrm{kin}} \approx \hbar\omega_{0} - I_{p} \approx 0.346~\text{a.u.},
\label{ener_kin}
\ee
as shown by the red curve in Fig.~\ref{fig1}(e).

Notably, for the specific case of CEP $\phi=0$, the central peak in the two-pulse energy spectrum appears at the same kinetic energy as in the single-pulse result. This makes the $\phi=0$ case a convenient reference for normalization and comparison, allowing CEP-induced redistributions of photoelectron yield to be identified relative to an unchanged peak position.

We note that the photoelectron momentum distribution obtained from the $2p_{-}$ ($m=-1$) initial state is the mirror image of that for the $2p_{+}$ ($m=+1$) state. This follows directly from the azimuthal phase structure of the initial bound state, $Y_{1m}(\theta,\varphi)\propto \Exp{im\varphi}$,where $\varphi$ denotes azimuthal angle. Reversing the sign of $m$ corresponds to complex conjugation of the angular dependence, which is equivalent to the spatial transformation $\varphi\rightarrow -\varphi$, or equivalently $p_y\rightarrow -p_y$, in the polarization plane. Since the linearly polarized driving field preserves reflection symmetry, all interference features of the PMD are retained under this transformation, with only the handedness reversed. Consequently, the momentum distributions for $m=\pm 1$ are related by mirror reflection and contain identical physical information.
  
As evident from Fig.~\ref{fig1}, variation of the CEP modifies the energy spectrum through quantum interference between wave packets released by the two delayed pulses. To further elucidate the combined role of CEP and interpulse delay, in Fig.~\ref{fig2} we present the integrated energy spectra for different values of these parameters.
  
\subsection{Effect of CEP and interpulse delay on PMD} \label{subsec:cep}

\begin{figure}[t]
\centering
\includegraphics[width=\linewidth]{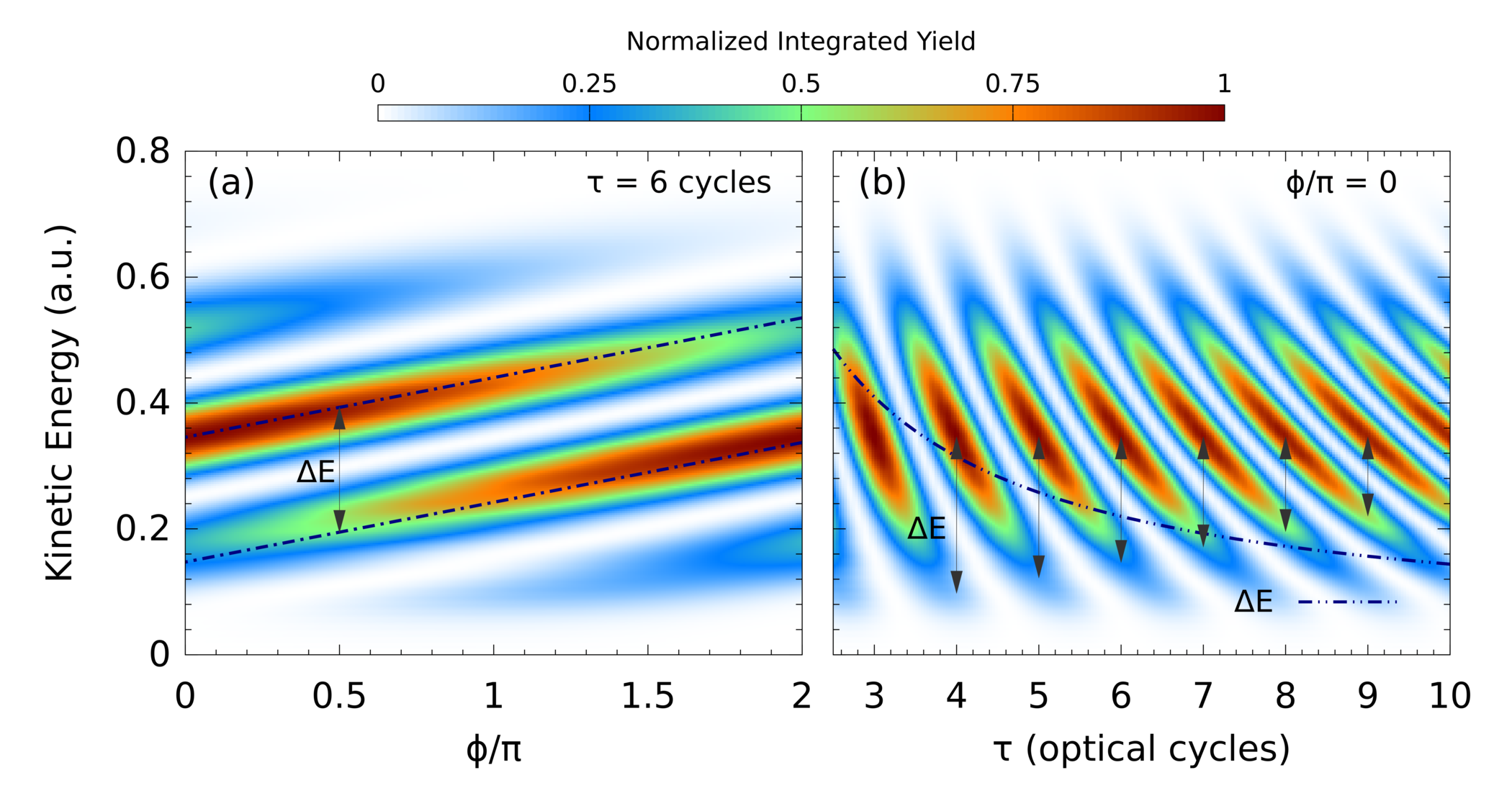}
\caption{Angle-integrated photoelectron energy yield as a function of (a) CEP $\phi$ at fixed interpulse delay $\tau=6$ optical cycles and (b) interpulse delay $\tau$ at fixed CEP $\phi=0$, for two linearly polarized XUV pulses. The bright ridges correspond to constructive interference between electron wave packets released by the two pulses. Arrows mark the energy spacing $\Delta E$ between adjacent interference fringes. The linear CEP dependence at fixed delay in (a) and the inverse scaling of the fringe spacing with delay, $\Delta E \propto 1/\tau$, in (b) are represented by dashed lines.}
\label{fig2}
\end{figure}

In Fig.~\ref{fig2}, the angle-integrated photoelectron energy yield as a function of the relative CEP and the interpulse delay is presented. In the single-photon XUV regime, ionization proceeds perturbatively, and the total photoelectron amplitude for a given kinetic energy $E$ can be written as the coherent sum of contributions from the two temporally separated pulses,
\be
A(E) = A_{1}(E) \Exp{-i \phi_1} + A_{2}(E) \Exp{-i \phi_2} \Exp{i E \tau},
\ee
however, in our case $\phi_1 = 0$ and $\phi = \phi_2 - \phi_1 = \phi_2$ is the relative CEP of the two pulses. So, the above 
equation modifies to,
\be A(E) = A_{1}(E)  + A_{2}(E) \Exp{i (E \tau - \phi)},\ee 
where $A_{1}(E)$ and $A_{2}(E)$ are the complex ionization amplitudes associated with the first and second pulses, respectively.
The corresponding photoelectron yield is
\be
|A(E)|^{2}
= |A_{1}(E)|^{2} + |A_{2}(E)|^{2}
+ 2\,\mathrm{Re}\!\left[A_{1}^{*}(E)A_{2}(E)\,\Exp{i(E\tau-\phi)}\right].
\ee
Since the individual contributions $|A_{1}(E)|^2$ and $|A_{2}(E)|^2$ are independent of the relative CEP and vary only slowly with energy, they form a smooth background. Consequently, all CEP- and delay-dependent modulations observed in Fig.~\ref{fig2} originate from the interference term.
 
The interference term oscillates with the accumulated Ramsey phase
\(E\tau-\phi\), which determines the positions of maxima and minima in the
photoelectron energy spectrum. The interference condition can be written as
\begin{equation}
	E\tau-\phi = n\pi ,
\end{equation}
with integer \(n\). Even values of \(n\) (\(n=2k\)) correspond to constructive
interference, leading to an enhanced photoelectron yield, whereas odd values
(\(n=2k+1\)) correspond to destructive interference, leading to a suppressed
photoelectron yield. These phase-matching conditions define the bright and dark
fringes observed in Fig.~\ref{fig2}.

In Fig.~\ref{fig2}(a), the interpulse delay is fixed at $\tau=6$ optical cycles and the CEP is varied. According to the above condition, the energies of the interference maxima depend linearly on $\phi$,
$E_n(\phi)=(2\pi n+\phi)/\tau$, giving rise to the straight, slanted ridges seen in the figure. The
dashed lines represent fits based on this linear dependence and show excellent agreement with the
numerical data.

In Fig.~\ref{fig2}(b), the CEP is fixed to $\phi=0$ and the interpulse delay $\tau$ is varied. In this
case, the same Ramsey condition predicts that the spacing between adjacent interference maxima scales
as
\be
\Delta E = \frac{2\pi}{\tau}.
\label{ramsey_condition}
\ee
This inverse dependence is directly highlighted by the arrows marking the fringe spacing in
Fig.~\ref{fig2}(b). The systematic reduction of $\Delta E$ with increasing delay confirms that the
energy-domain modulations originate from phase accumulation between the two temporally separated
ionization pathways, fully consistent with the Ramsey-spectroscopy framework.

\subsection{Influence of additional laser parameters} \label{subsec:params}

\begin{figure}[b]
\centering
\includegraphics[width=\linewidth]{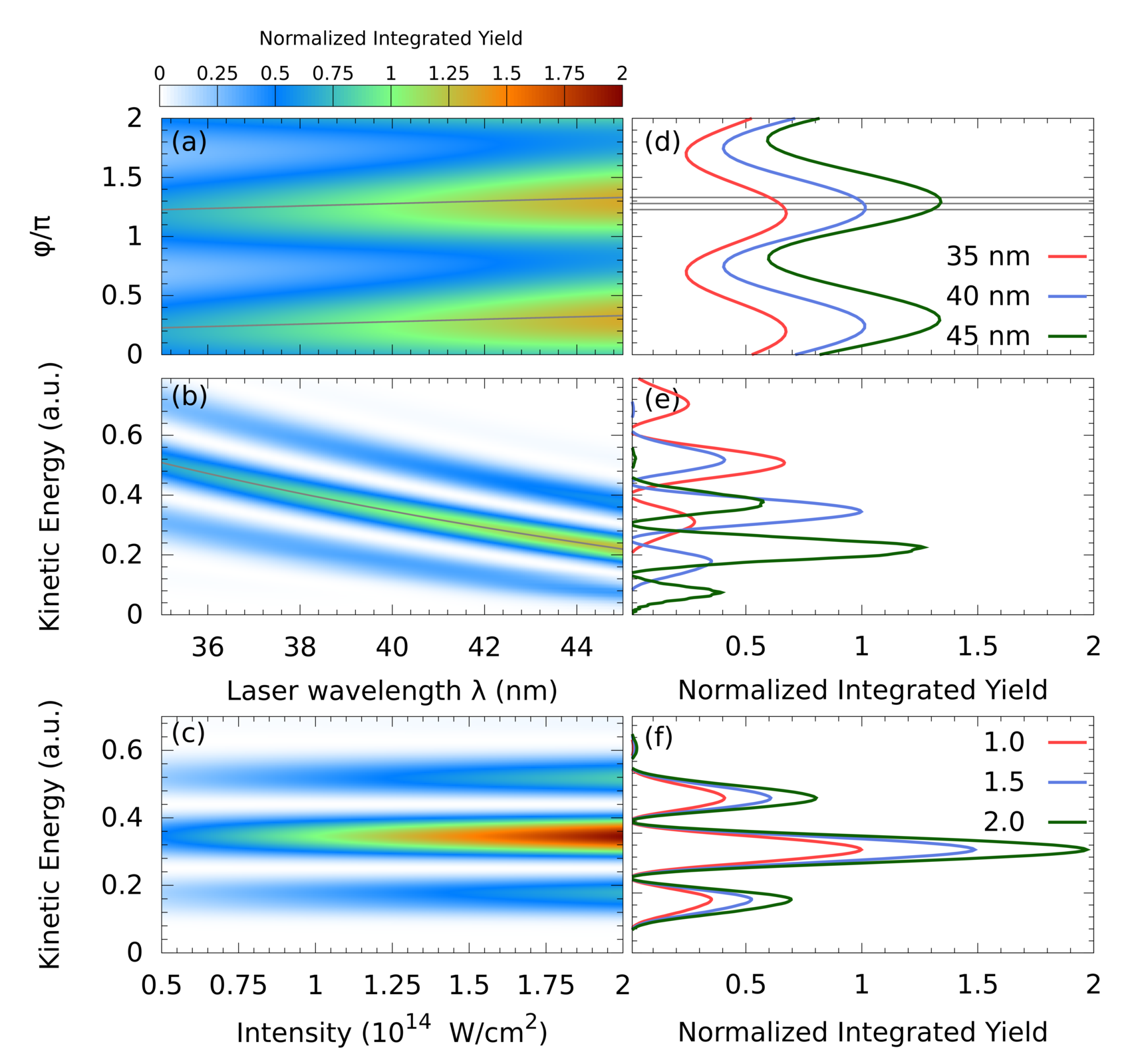}
\caption{Effect of laser wavelength and intensity on the photoelectron spectra for fixed CEP
$\phi=0$ and interpulse delay $\tau=6$ optical cycles.
(a) Energy-integrated angular distribution as a function of laser wavelength $\lambda$.
(b) Corresponding angle-integrated energy--wavelength map showing a systematic shift of the
interference fringes with photon energy.
(c) Energy--intensity map with all other laser parameters fixed.
Panels (d)--(f) show representative spectra extracted from panels (a)--(c), respectively.
In panel (f), the intensity is expressed in units of $10^{14}$~W/cm$^2$.}
\label{fig3}
\end{figure}

To further examine the robustness of the interference patterns, we investigate the dependence
of the photoelectron spectra on additional laser parameters, namely the laser wavelength and peak intensity. The wavelength is varied from 35 to 45~nm, while the intensity is scanned
from $0.5\times10^{14}$ to $2\times10^{14}$~W/cm$^2$. Throughout this analysis, the CEP is fixed at $\phi=0$, the pulse duration is held at $5\tau_0$, and the interpulse delay is fixed at
$6\tau_0$, where $\tau_0=2\pi/\omega_0$ denotes the optical cycle corresponding to the respective wavelength. The results are summarized in Fig.~\ref{fig3}.

The energy-integrated angular distribution as a function of wavelength is shown in
Fig.~\ref{fig3}(a) for a fixed intensity of $10^{14}$~W/cm$^2$. A weak but systematic shift of the
angular distribution with increasing wavelength is observed. This behavior can be understood
from the kinematic relation $\tan\varphi = p_{\perp}/p'_{\parallel}$, where $p_{\perp}$ and $p'_{\parallel}$ are the transverse and scaled longitudinal components of the asymptotic photoelectron momentum. Since the number of optical cycles is kept fixed, the pulse duration scales linearly with wavelength, $T = 2\pi N/\omega_0 \propto \lambda$, leading to an effective dilation of the longitudinal momentum scale with increasing $\lambda$. To compare different wavelengths on a common footing, the longitudinal momentum is therefore rescaled according to
$p'_{\parallel} = p_{\parallel}(\lambda_0/\lambda)$, with $\lambda_0=40$~nm chosen as the reference wavelength. Using this scaled momentum, the ridge in Fig.~\ref{fig3}(a) follows a simple tangent law
\be
\varphi \approx \arctan\!\left(\frac{\lambda}{\lambda_0}
\frac{1}{\sqrt{2E_{\mathrm{kin},\lambda}}}\right).
\ee

In the single-photon XUV regime, the kinetic energy is given by
\be
E_{\mathrm{kin},\lambda} \approx E_{\mathrm{ph}}(\lambda) - I_p ,
\label{kin0}
\ee
where $E_{\mathrm{ph}}(\lambda)$ denotes the photon energy associated with the wavelength
$\lambda$. The ponderomotive energy $U_p$ is negligibly small at the intensities considered here
and is therefore omitted.

The corresponding angle-integrated energy--wavelength map is shown in Fig.~\ref{fig3}(b). The
interference fringes shift systematically with wavelength, reflecting the change in photon
energy. This behavior is expected for single-photon XUV ionization, where the absolute fringe
positions are governed by Eq.~\eqref{kin0}, while the fringe spacing is fixed by the interpulse
delay through the Ramsey phase condition [Eq.~\eqref{ramsey_condition}]. The representative
spectra in Fig.~\ref{fig3}(e) further demonstrate that, although the peak positions shift with
wavelength, the overall interference structure is preserved.

In contrast, Fig.~\ref{fig3}(c) illustrates the effect of varying the laser intensity at a fixed
wavelength. The interference fringes persist with an essentially constant energy separation as
the intensity is increased, as highlighted by the spectra in Fig.~\ref{fig3}(f). While the total
yield increases proportionally with intensity—reflecting the larger number of photons involved
in the ionization process—the fringe spacing remains unchanged. This behavior is incompatible
with an Autler--Townes splitting mechanism \cite{yang2025}, which would lead to an
intensity-dependent splitting proportional to the Rabi frequency. The absence of such a
dependence therefore rules out Autler--Townes physics.

Instead, the observed intensity-independent fringe separation provides compelling evidence that
the spectral modulations originate from Ramsey-type temporal interference between electron wavepackets released by the two time-delayed pulses. In this picture, the fringe spacing is determined solely by the interpulse delay [Eq.~\eqref{ramsey_condition}] and is insensitive to both the laser wavelength (apart from trivial energy shifts) and the field strength. On the basis of these observations, all subsequent results are presented for a wavelength of 40~nm, pulse durations of $5\tau_0$, an interpulse delay of $6\tau_0$, and a peak intensity of $10^{14}$~W/cm$^2$, unless  otherwise stated.

\subsection{Population dynamics} \label{subsec:population}

\begin{figure}[t]
\centering
\includegraphics[width=\linewidth]{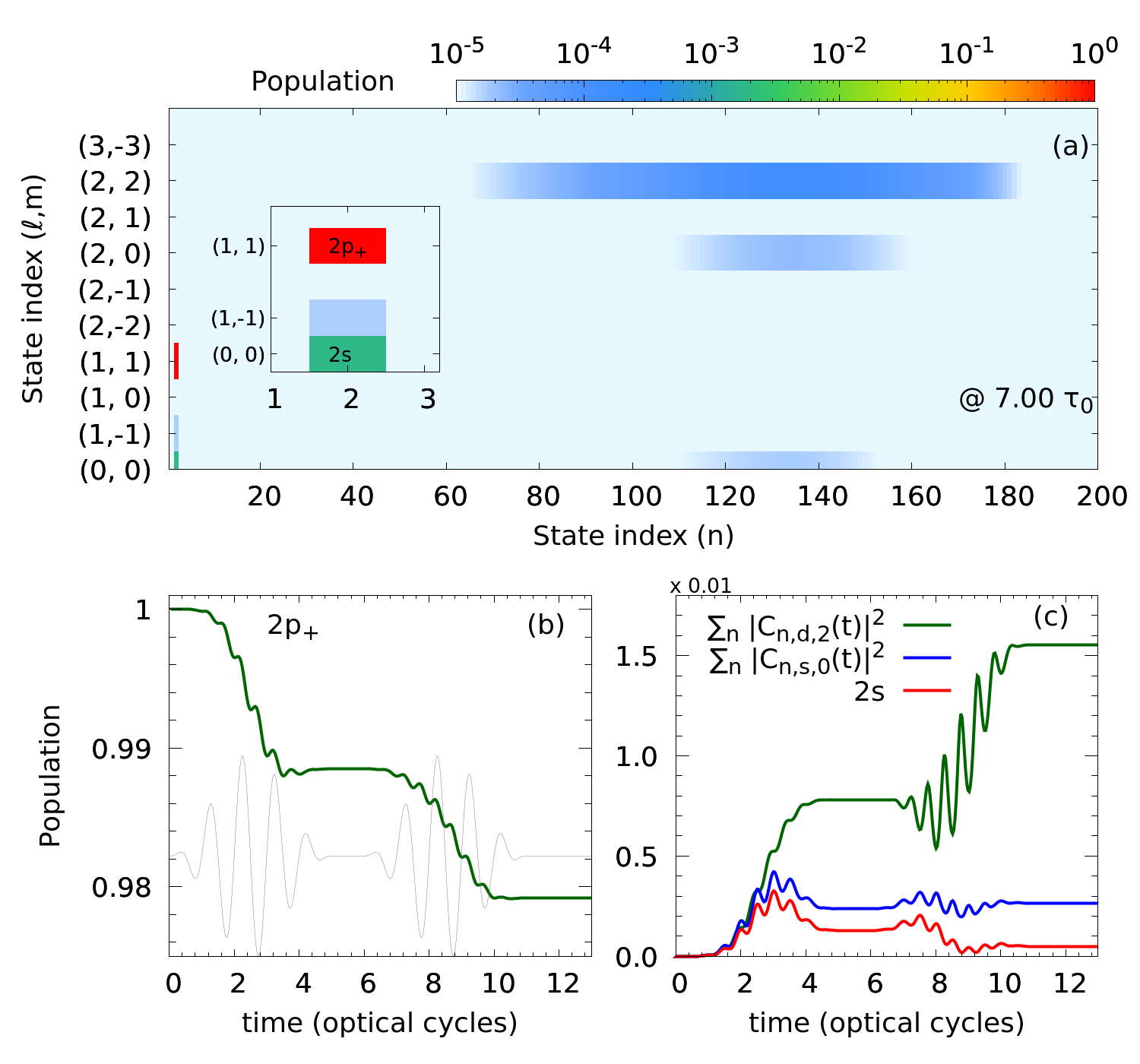}
\caption{Time-resolved population dynamics illustrating the role of CEP in the formation of
interference patterns.
(a) Snapshot of the population distribution over field-free eigenstates $(n,\ell,m)$ at
$t=7\tau_0$, showing population transfer from the initially prepared $2p_{+}$ state into
nearby bound and continuum states.
(b) Temporal evolution of the $2p_{+}$ population, demonstrating stepwise depletion during
the interaction with the two time-delayed laser pulses; the schematic laser pulse envelope
is shown for reference.
(c) Time-resolved population summed over different angular-momentum channels, together with
the population of the $2s$ state, highlighting transient population trapping and redistribution
during the delay interval.}
\label{fig4}
\end{figure}

To gain microscopic insight into how the carrier-envelope phase (CEP) controls the observed
interference patterns in the photoelectron spectra, we analyze the time-resolved population
dynamics during the interaction with the two delayed laser pulses. 

As discussed previously, the time-dependent wavefunction obtained from the TDSE
propagation is expanded in the field-free eigenbasis according to Eq. \eqref{psi_t}, 
and hence $|C_{n,\ell,m}(t)|^2$ represents the probability of the wavefunction $\psi(\mb{r},t)$ to be in the state $(n,\ell,m) \equiv R_{nl}(r) Y_{\ell m}(\theta,\varphi)$ at a given time $t$. It is evident that the laser-driven dynamics involves population transfer among a large number of bound and continuum states with different $(n,\ell,m)$ quantum numbers. The resulting evolution therefore represents an intricate coherent interplay of multiple excitation and ionization pathways. As a representative case, we present a time-snapshot of the population of all the levels at $t = 7\tau_0$ in Fig. \ref{fig4}(a).   

It can be seen from this figure that, starting from the initially prepared current-carrying $2p_{+}$ state, population is transferred not only to continuum states but also to nearby bound states, 
most notably the $2s$ state. This population redistribution provides the microscopic origin of the multiple interfering ionization pathways responsible for the energy-domain fringes observed in the PMD.

\begin{figure}[b]
\centering
\includegraphics[width=\linewidth]{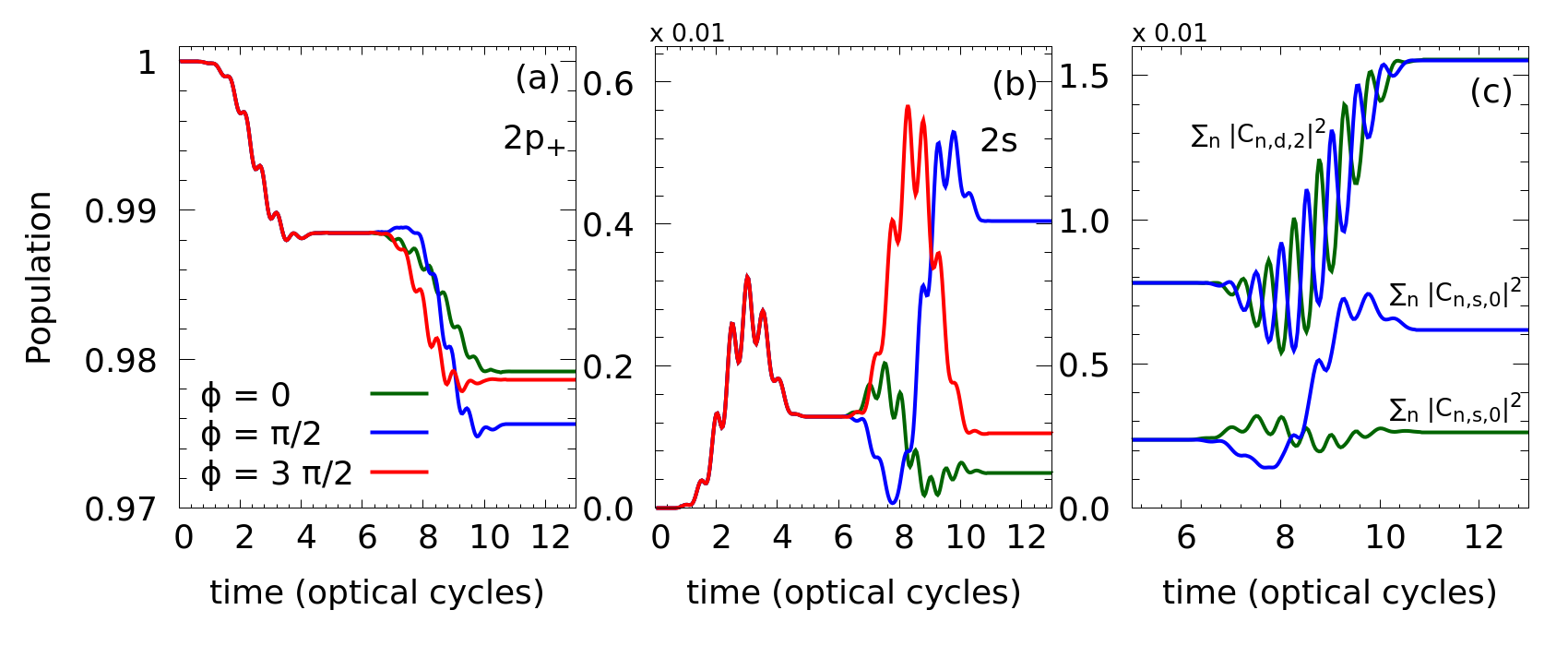}
\caption{CEP-dependent population dynamics for fixed interpulse delay $\tau=6\tau_0$.
(a) Depletion of the initially populated $2p_{+}$ state for different CEP values.
(b) Time-resolved population of the $2s$ state, showing strong sensitivity to the CEP.
(c) Population summed over different principal quantum numbers and angular-momentum channels,
illustrating CEP-controlled redistribution among bound and continuum states.
All quantities are shown as a function of time in units of optical cycles.}
\label{fig5}
\end{figure}

The temporal depletion of the $2p_{+}$ state is shown in Fig.~\ref{fig4}(b). The step-like
reduction of the population reflects the sequential interaction with the first and second
pulses, as emphasized by the overlaid schematic of the laser field. Importantly, the
population does not decay monotonically but exhibits plateaus during the delay interval,
indicating coherent superpositions of bound and continuum components that persist during the interpulse delay.

From a presentation perspective, it is neither practical nor particularly illuminating to identify a single, well-defined ionization pathway, since the observed interference effects emerge from the collective superposition of many such channels. To obtain a more transparent physical picture, Fig.~\ref{fig4}(c) presents the population summed over all $(s,0)$ and $(d,2)$ channels, $\sum_{n}|C_{n,s,0}(t)|^{2}$ and $\sum_{n}|C_{n,d,2}(t)|^{2}$ respectively along with the explicit population of the $2s$ state as well. This shows that the $2s$ level provides the dominant contribution to the summed $(s,0)$ channel population during the interaction and the subsequent delay interval. Motivated by this observation, we therefore focus in the following on the time-resolved population dynamics of the $2s$ state as a representative indicator of CEP-controlled bound-state participation in the interference process. It can be seen that the $2s$ population shows pronounced temporal modulation, demonstrating that transient population trapping and bound–bound coupling play an important role in shaping the subsequent ionization by the second pulse.

Furthermore, the influence of the CEP on these dynamics is highlighted in Fig.~\ref{fig5}, where the population evolution is compared for different CEP values. While the depletion of the
initial $2p_{+}$ state [Fig.~\ref{fig5}(a)] is only weakly affected by the CEP, the population
of the $2s$ state [Fig.~\ref{fig5}(b)] exhibits a pronounced CEP dependence. This sensitivity
indicates that the CEP primarily controls the relative phase between different excitation pathways rather than the total ionization probability. In Fig.~\ref{fig5}(c), the summed population for the $(d,2)$ and $(s,0)$ channels are also compared for $\phi = 0$ and $\pi/2$, and it can be seen that the $(d,2)$ channels primarily exhibit a relative phase shift in their population dynamics, however, the $(s,0)$ channels clearly show distinct temporal evolution. The summed population shown in Fig.~\ref{fig5}(c) confirms that the CEP redistributes population among bound and continuum channels without significantly altering the overall depletion.

\begin{figure}[t]
\centering
\includegraphics[width=\linewidth]{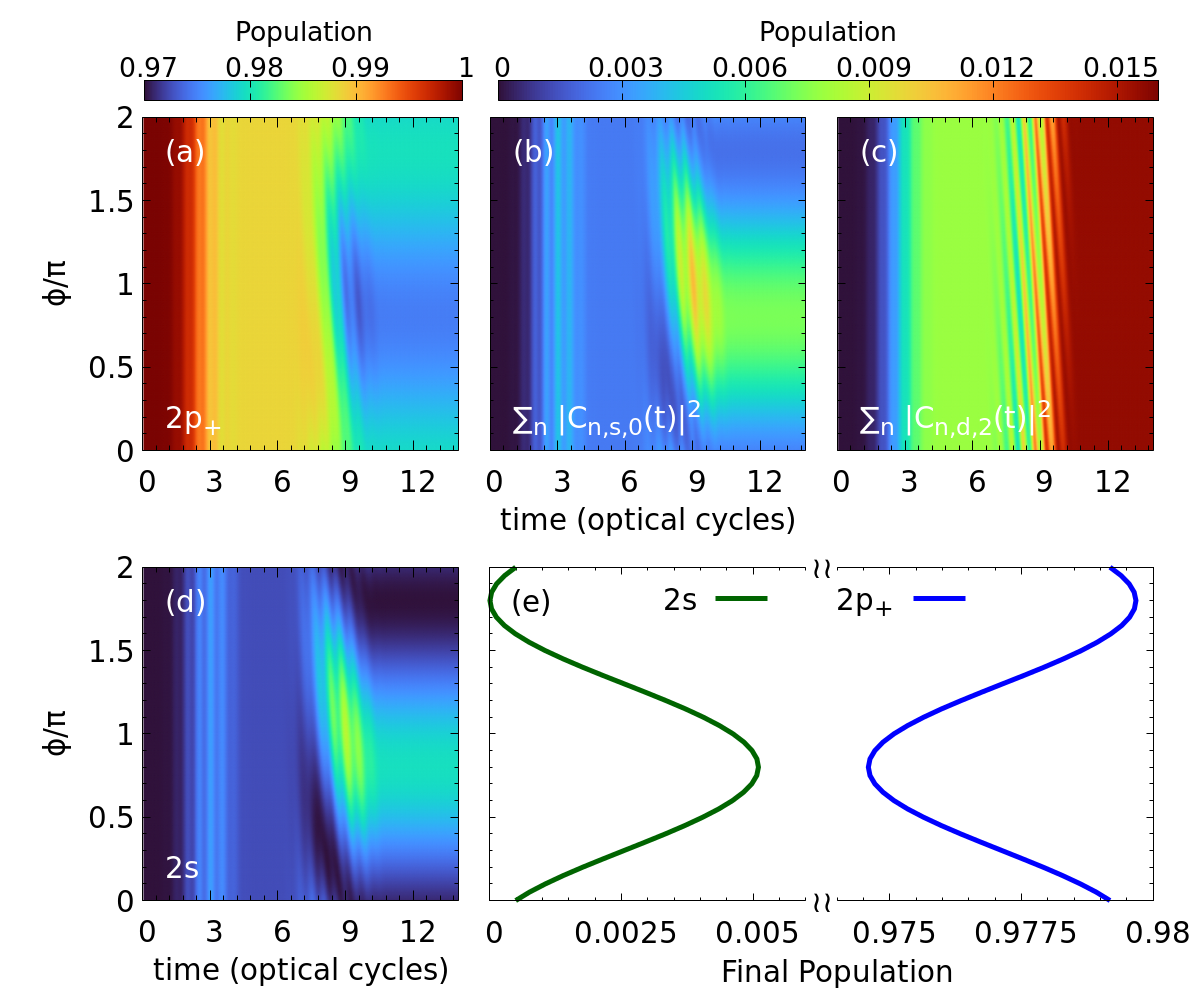}
\caption{The CEP-temporal map of the population is presented for initial state $2p_{+}$ (a), along with
the summed $(s,0)$ (b) and $(d,2)$ (c) channels, however, it is explicitly presented for $2s$ state in (d). 
The final population (after the second pulse) as a function of CEP  for $2s$ and $2p_{+}$ is also shown in (e).}
\label{fig6}
\end{figure}

While such line plots provide clear insight into the dynamics for selected phases, they do not fully capture the global phase dependence of the population redistribution. To obtain a comprehensive view, Fig.~\ref{fig6} presents CEP--time maps of the time-resolved population dynamics for the initial $2p_{+}$  state and selected bound-state manifolds. This representation reveals the continuous evolution of the population with CEP and highlights systematic phase shifts of the population modulations induced by the second pulse. In particular, the CEP--temporal maps make explicit the Ramsey-type interference between excitation pathways launched by the two pulses, establishing a direct connection between CEP-controlled bound-state dynamics and the interference structures observed in the photoelectron spectra.

Panel~(a) of Fig.~\ref{fig6} shows the CEP--time map of the population remaining in the initial current-carrying $2p_{+}$  state, demonstrating a smooth yet phase-sensitive depletion during the interaction with the two time-delayed XUV pulses. The corresponding population transferred into the $s$-wave ($\ell=0,m=0$) and $d$-wave ($\ell=2,m=2$) manifolds, summed over the principal quantum number $n$, is shown in panels~(b) and (c), respectively. These maps exhibit pronounced CEP-dependent modulations synchronized with the temporal overlap of the second pulse, indicating interference between excitation pathways initiated by the two pulses.

To highlight the role of a specific bound state, panel~(d) explicitly displays the CEP–time map
of the $2s$ population. Compared to the summed $s$-channel population in panel~(b), the $2s$
state captures the dominant CEP-dependent modulation, confirming that it provides the primary
contribution to the low-lying bound-state dynamics in the present parameter regime. The strong
phase sensitivity observed during the action of the second pulse reflects the accumulation of
a relative phase between the excitation amplitudes associated with the two pulses, in direct
analogy with Ramsey-type interference.

Finally, panel~(e) shows the final populations of the $2s$ and $2p_{+}$ states after the end of
the second pulse as a function of CEP. The anti-correlated CEP dependence of these populations
demonstrates that the carrier–envelope phase does not merely influence transient dynamics, but
provides a robust control knob for steering population redistribution among bound states.
Together, Fig.~\ref{fig6} establishes a clear link between CEP-controlled phase accumulation,
time-resolved bound-state dynamics, and the interference structures observed in the
photoelectron momentum and energy distributions.

These observations provide a clear physical picture for the CEP control of the interference
patterns discussed earlier. By modifying the relative phase accumulated between population
components created by the first pulse and those driven by the second pulse, the CEP controls
the coherent superposition of ionization pathways. This CEP-dependent population redistribution,
particularly involving the $2s$ state, ultimately manifests as the modulation of interference
fringes in the photoelectron momentum and energy distributions.

\subsection{CEP enabled Dynamic Stark shift} \label{subsec:stark}

\begin{figure}[t]
\centering
\includegraphics[width=\linewidth]{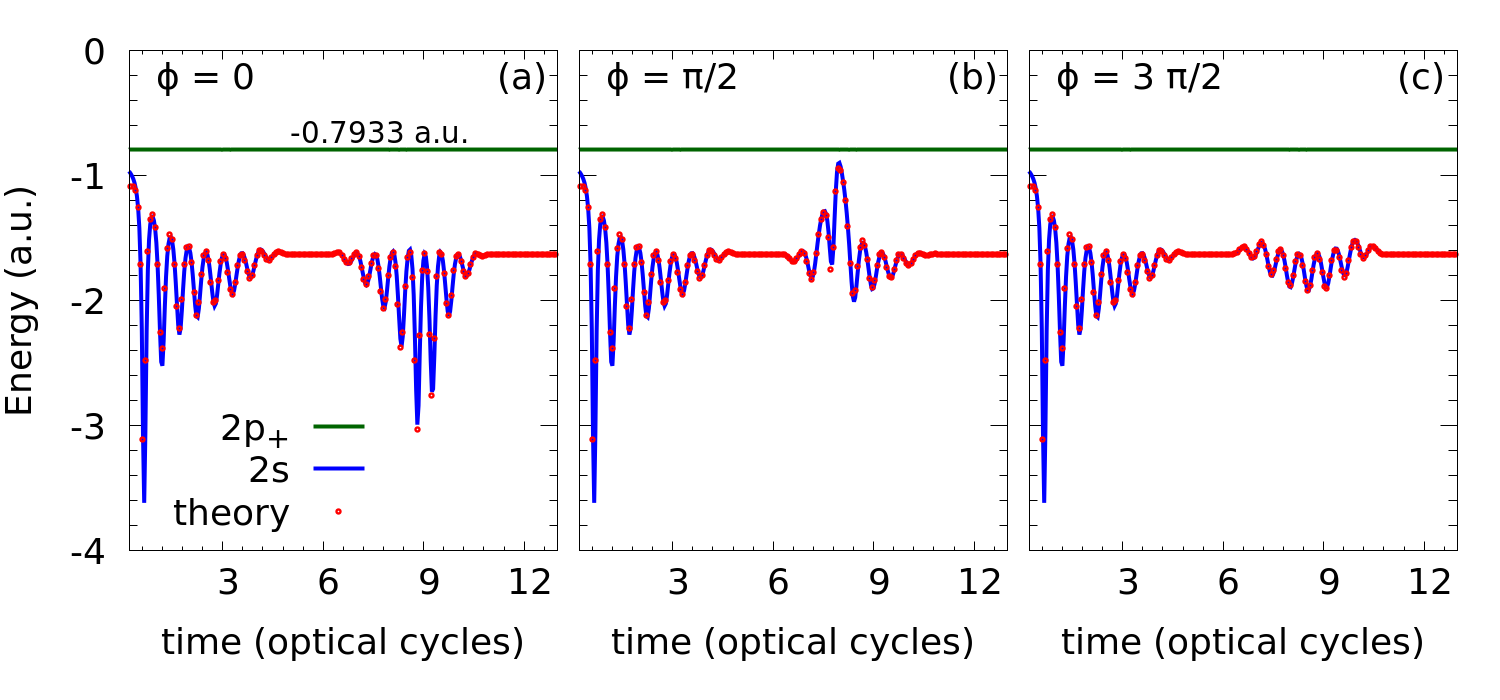}
\caption{Time evolution of the dressed energy level of $2s$ for CEP $\phi = 0$ (a), $\phi = \pi/2$ (b) and $\phi = 3\pi/2$ (c) are presented, Eq. \eqref{dressed_ener}. The energy level of the $2p_{+}$ is also shown for reference and exhibits negligible temporal dependence, remaining effectively constant on the scale of the figure for different CEP values. The theoretical estimates from reduced two level model as given by Eq. \eqref{e2s_time} are plotted as red dots for all the CEP cases.}
\label{fig7}
\end{figure}

As we observed that the CEP controls the bound-bound coupling dynamics through the dominant 
$2s$ channel. This CEP control can be understood in the context of the phase accumulated 
after the interaction of the second pulse.This phase accumulation can be attributed to the dynamic Stark shift of the $2s$ state, i.e., the energy shift of the field-dressed $2s$ state. As the TDSE wave function is expanded in the energy eigenbasis of the field-free Hamiltonian, each expansion coefficient naturally acquires a time-dependent phase determined by the corresponding eigenenergy. Accordingly, the coefficient  $C_{2,0,0}(t) \equiv C_{2s}(t) = |C_{2s}(t)| \Exp{-i \int^t E_{2s}(t') dt'}$, and similarly $C_{2,1,1}(t) \equiv C_{2p_+}(t) $ can also be expressed in polar form as well with appropriate phase. The dressed energy of the $2s$ and $2p_+$ levels can be written as:
\be E_{nlm}(t) = -\frac{d}{dt} \arg\!\left[C_{n,\ell,m}(t)\right]  . \label{dressed_ener}\ee
In Fig. \ref{fig7} (solid lines), we have presented the time-dependent dressed energy level of $2s$ state together with the energy level of the $2p_+$ state (which shows negligible temporal dependent variation) as given in Eq. \eqref{dressed_ener} for different CEP values. It should be noted that the $C_{2s}(t)$ in Eq. \eqref{dressed_ener} is obtained through full TDSE solver without any approximation, wherein the wavefunction is given by Eq. \eqref{psi_t}. These variations of the energy level with CEP is very well reproduced by the reduced two level model. It can be understood that the CEP controls the interaction between the two bound states $2p_{+} \leftrightarrow 2s$ as also apparent from the Fig. \ref{fig6}(e), wherein the final population of the $2p_+$ and $2s$ states are anti-correlated, i.e. if one is decreasing then other is increasing and vice-versa. We can approximate the wavefunction as two level system, $\ket{\psi(\mb{r},t)} \approx C_{2s}(t) \ket{2s} + C_{2p_+}(t) \ket{2p_+}$, and the dressed energy of the $2s$ level is obtained as Eq. \eqref{app_E2s}:   
\be
E_{2s}(t) = E_{2s}^0 - \frac{|d_x|\,E_x(t)}{|C_{2s}(t)|^2}
\Re\!\left[C_{2s}^*(t)C_{2p_{+}}(t)\right],
\label{e2s_time}
\ee
where, $|d_x| \approx 0.37$ is the dipole matrix transition element obtained in this SAE based calculations and $E_{2s}^0 = -1.629$ a.u. is the eigenvalue of the field free Hamiltonian, i.e. $H_0 \ket{2s} = E_{2s}^0 \ket{2s}$, and $E_x(t)$ is the time-dependent laser field. The time dependent dressed energy of the $2s$ state given by Eq. \eqref{e2s_time} is also presented in the Fig. \ref{fig7} along with the directly obtained from full TDSE Eq. \eqref{dressed_ener}, and this reduced two-level model quantitatively captures the CEP-enabled dynamic Stark shift, which eventually affects the ionization pathway in terms of the acquired phase which manifests in the Ramsey interference of the ionization pathways generated by two pulses.    

The quantitative agreement between the dressed energy extracted directly from the TDSE [Eq. \eqref{dressed_ener}] and the reduced two-level expression [Eq. \eqref{e2s_time}] can be understood from the structure of the phase evolution. 
%The instantaneous phase velocity of a TDSE expansion coefficient is determined exclusively by coherent coupling to states with non-negligible amplitude and dipole connectivity. 
As demonstrated by the time-resolved population analysis, during the interaction the bound-state dynamics are dominated by the $2p_{+}$ and $2s$ states, while population in other bound states remains negligible. Moreover, for the chosen polarization, the laser field couples $2p_{+}$ and $2s$ directly, whereas dipole coupling to other bound states is either forbidden or dynamically suppressed.

Although coupling to the continuum leads to depletion of the bound-state amplitudes, it does not contribute coherently to the phase derivative $-d\,\arg[C_{2s}(t)]/dt$, since continuum components rapidly dephase and do not maintain a fixed phase relation with the bound states. Consequently, the rate change of phase of the $2s$ probability amplitude is governed entirely by its coherent interaction with the $2p_{+}$ state. Under these conditions, projecting the TDSE onto the $\{2s,2p_{+}\}$ subspace captures all contributions relevant to the phase evolution, explaining why the reduced two-level model reproduces the TDSE-derived dressed energy of the $2s$ level.

\section{Conclusion} \label{sec:conclusion}

We have shown that two time-delayed XUV pulses can induce Ramsey-type quantum interference in photoelectron momentum and energy distributions in single-photon ionization. Full-dimensional TDSE simulations for neon initially prepared in a current-carrying $2p_+$ state demonstrate that the observed interference fringes arise from the coherent superposition of electron wave packets released by the two pulses. The fringe positions obey a simple phase-matching condition determined by the interpulse delay and the relative carrier-envelope phase, resulting in a linear dependence on the CEP and an inverse scaling of the fringe spacing with the interpulse delay.

By varying the laser wavelength and intensity, we establish that the fringe spacing is insensitive to field strength, thereby excluding Autler--Townes splitting as the underlying mechanism. Instead, the interference is governed by temporal phase accumulation between distinct ionization pathways, consistent with the Ramsey interference picture. Time-resolved population dynamics reveal that the carrier-envelope phase primarily controls the relative phase and transient population of low-lying bound states, with the $2s$ state providing the dominant contribution to the bound-state dynamics. The CEP-dependent phase accumulation can be interpreted as a dynamic Stark shift of the dressed bound states, which is accurately captured by a reduced two-level description.

These results provide a unified description of CEP-controlled temporal interference in XUV photoionization and clarify the role of bound-state dynamics in shaping continuum interference patterns.

\section*{Acknowledgments} 
The authors acknowledge the Department of Science
and Technology (DST) for providing computational resources through the FIST program (Project No. SR/FST/PS-
1/2017/30). Also authors acknowledge BITS - Pilani, Pilani Campus for providing HPC facility. 

\section*{Data Availability}

The data that support the findings of this article are not
publicly available. The data are available from the authors
upon reasonable request.
 
\appendix
\section{Dressed energy of the $2s$ state} \label{app:2s}

To elucidate the origin of the instantaneous dressed energy of the $2s$ state used in the main
text, we briefly outline the derivation within a reduced two-level description. Restricting the
wavefunction to the dominant bound-state subspace, we write
\be
\ket{\psi(\mb{r},t)} = C_{2s}(t)\ket{2s} + C_{2p_{+}}(t)\ket{2p_{+}} .
\ee
Solving the TDSE in the length gauge and projecting onto the $2s$ state yields
\be
i\dot C_{2s}(t) = E_{2s}^0 C_{2s}(t) - d_x\,E(t)\,C_{2p_{+}}(t),
\ee
where
\be
d_x \equiv \langle 2s | r\sin\theta\cos\phi | 2p_{+} \rangle
\ee
is the dipole transition matrix element along the laser polarization direction, which for our SAE potential is calculated to be $|d_x| \sim 0.37$. We can express the complex amplitude in polar form,
\(
C_{2s}(t)=|C_{2s}(t)|\Exp{-i \int^t E_{2s}(t') dt'},
\)
the instantaneous energy associated with the $2s$ component is defined as:
\be
E_{2s}(t)\equiv  -\frac{d}{dt} \arg\!\left[C_{2s}(t)\right] 
= \Re\!\left[\frac{i\dot C_{2s}(t)}{C_{2s}(t)}\right].
\ee
Substituting the equation of motion for $\dot C_{2s}(t)$ and noting that $E_{2s}$ is real, we obtain
\be
E_{2s}(t) = E_{2s}^0 - |d_x|\,E_x(t)\, \Re\!\left[\frac{C_{2p_{+}}(t)}{C_{2s}(t)}\right].
\ee
Finally, rewriting the real part explicitly leads to
\be
E_{2s}(t) = E_{2s}^0 - \frac{|d_x|\,E_x(t)}{|C_{2s}(t)|^2} \Re\!\left[C_{2s}^*(t)C_{2p_{+}}(t)\right],
\label{app_E2s}
\ee
This quantity represents the instantaneous rate change of phase of the $2s$ probability amplitude and captures the CEP-dependent laser dressing arising from coherent dipole coupling to the $2p_{+}$ state.

\bibliographystyle{apsrev4-2}

%\bibliography{Bibliography}

\end{document}